\documentclass[conference]{IEEEtran}
\IEEEoverridecommandlockouts

\usepackage{cite}
\usepackage{amsmath,amssymb,amsfonts}

\usepackage{graphicx}
\usepackage{booktabs}
\usepackage{makecell}

\usepackage{textcomp}
\usepackage{xcolor}

\usepackage{caption}
\usepackage{subcaption}

\usepackage{tcolorbox}

\usepackage[hidelinks]{hyperref}

\usepackage{algpseudocode,algorithm,algorithmicx}
\usepackage{ifthen}

\usepackage[frozencache,cachedir=.]{minted}

\usepackage{color}
\usepackage{alltt, listings, textcomp, verbatim}

\algnewcommand\algorithmicforeach{\textbf{for each}}
\algdef{S}[FOR]{ForEach}[1]{\algorithmicforeach\ #1\ \algorithmicdo}

\definecolor{darkred}{rgb} {0.5,0.0,0.0}
\definecolor{darkblue}{rgb} {0.0,0.0,0.5}
\definecolor{darkgreen}{rgb} {0.0,100.0,0.0}

\newcommand{\john}   [1]{{{\color{darkblue}(john) #1}}}
\newcommand{\osama}   [1]{{{\color{darkred}(osama) #1}}}
\newcommand{\serban}   [1]{{{\color{darkgreen}(serban) #1}}}

\newcommand{\final}{1} 
\ifthenelse{\equal{\final}{1}}{\renewcommand{\john}[1]{}}{}
\ifthenelse{\equal{\final}{1}}{\renewcommand{\osama}[1]{}}{}
\ifthenelse{\equal{\final}{1}}{\renewcommand{\serban}[1]{}}{}

\def\BibTeX{{\rm B\kern-.05em{\sc i\kern-.025em b}\kern-.08em
    T\kern-.1667em\lower.7ex\hbox{E}\kern-.125emX}}
\begin{document}

\title{Essentials of Parallel Graph Analytics}

\author{\IEEEauthorblockN{Muhammad Osama}
    \IEEEauthorblockA{University of California, Davis \\
        mosama@ucdavis.edu}
    \and
    \IEEEauthorblockN{Serban D. Porumbescu}
    \IEEEauthorblockA{University of California, Davis \\
        sdporumbescu@ucdavis.edu}
    \and
    \IEEEauthorblockN{John D. Owens}
    \IEEEauthorblockA{University of California, Davis\\
        jowens@ucdavis.edu}}

\maketitle

\begin{abstract}
We identify the graph data structure, frontiers, operators, an iterative loop structure, and convergence conditions as essential components of graph analytics systems based on the native-graph approach. Using these essential components, we propose an abstraction that captures all the significant programming models within graph analytics, such as bulk-synchronous, asynchronous, shared-memory, message-passing, and push vs.\ pull traversals. Finally, we demonstrate the power of our abstraction with an elegant modern C++ implementation of single-source shortest path and its required components.
\end{abstract}

\begin{IEEEkeywords}
    parallel, graph analytics, graph traversal, algorithms
\end{IEEEkeywords}


\section{Introduction}
\label{sec:introduction}

Graph analytics are used to study graphs and relationships between objects. Unfortunately, efficient analysis is increasingly difficult given increases in graph scale and workload diversity. Research in parallel graph analytics has tried to keep up by introducing new programming models, abstractions, and algorithms. Here, we explore one approach called \textbf{``native-graph'' graph analytics} and make the following contributions:

\begin{enumerate}
    \item We propose a framework, centered around an abstraction, that supports many design choices in each of the four pillars of graph analytics in the McCune et al.~\cite{McCune:2015:TLA} survey paper \emph{Thinking Like a Vertex (TLAV)}.
    \item An example, using modern C++ concepts, highlighting design choices and exercising our proposed abstraction.
\end{enumerate}

Our abstraction focuses on the graph as a ``native'' data structure, the active working set, and the ability to express iterative loops with convergence conditions to organize and schedule the computation and completion of a graph algorithm.

\section{Overview}
The native-graph approach for graph analytics focuses on operating directly on the graph's vertices or edges, or their relationships within the graph data structure, as opposed to linear-algebra-based graph analytics, which exploits the duality between graphs and sparse matrices to perform graph algorithms as sparse linear algebra operations. We focus on the native-graph approach and on implementing graph analytics with the different design choices described in TLAV~\cite{McCune:2015:TLA} while using a general abstraction that targets many of these design choices. TLAV describes the four pillars of graph analytics as deeply connected, interesting design decisions to be made in a graph framework implementation targeting specific hardware architectures. The pillars and relevant models we target are:

\begin{enumerate}
    \item \textbf{Timing}: Synchronous and asynchronous.
    \item \textbf{Communication}: Shared-memory and message-passing.
    \item \textbf{Execution Model}: Vertex programs and push vs.\ pull.
    \item \textbf{Partitioning}: Heuristics.
\end{enumerate}

We refer readers to the TLAV paper for an extensive discussion~\cite{McCune:2015:TLA} of these design choices. Many graph libraries carve a slice of the TLAV design space to implement graph algorithms; for example, Gunrock's~\cite{Wang:2017:GGG} and Pregel's~\cite{Malewicz:2010:PAS} bulk-synchronous programming model on a shared-memory system, or PowerGraph's~\cite{Gonzalez:2012:PDG} asynchronous and message-passing system. These libraries focus on a small subset of possible design decisions and are limited by abstractions that are tightly coupled to those design decisions. Although these systems make useful design decisions, we desire greater flexibility. We show that with an abstraction focused around a native graph data structure, essential components of graph analytics can target many of the models within TLAV in a single graph framework. A general abstraction that is able to capture more than a single slice of these four pillars allows for more versatility when expressing graph algorithms and targeting different hardware architectures, and also leaves room for scalability in the future. This paper is our vision of such an abstraction and how it is implemented.

\section{Abstraction for Each of the Four Pillars}
\label{sec:abstraction}
Our goal is to build a graph framework able to support many (or more) of the design choices within TLAV\@. We first describe the essential components:
\begin{enumerate}
    \item A \textbf{graph data structure} that expresses the underlying graph representation;
    \item \textbf{Frontiers}, active sets of vertices or edges in each iteration of a graph algorithm;
    \item \textbf{Operators}, programs operating on the graph data structure or the frontiers. Operators are often defined as traversals or transformations on the frontiers/graphs; and
    \item \textbf{Loop structure/convergence condition(s)} to organize and schedule the computation and completion of a graph algorithm.
\end{enumerate}

Before we show how our abstraction, built on the essential components above, targets the different programming models, we emphasize that these models are heavily interdependent, and describing how our abstraction targets them independently of each other will provide an incomplete understanding. Therefore, the following sections may cross-reference the models that are not strictly within the same ``pillar''. We also summarize the models we capture and ignore within our abstraction and framework in Table~\ref{tab:summary}.

\subsection{Synchrony and Asynchrony}
\label{sec:timing}

In the execution domain, a timing model is often the building block for a graph framework. The choice of a bulk-synchronous programming model implies that the computation is performed in bulk \emph{supersteps}, and a global synchronization barrier is used to synchronize the completion of each superstep. In contrast, asynchronous programming models have no explicitly defined barriers, and work is performed whenever the required resources are available~\cite{McCune:2015:TLA}. In the TLAV survey paper, asynchronous execution models are defined to be more ``complex'' than bulk-synchronous execution models but allow for better workload balance.

\textbf{The abstraction that targets the timing model in our framework is structuring operators with support for \emph{execution policies}.} We center our execution domain around operations on graphs or frontiers (an active vertex or edge set)~\cite{Wang:2017:GGG}, and how these operators are structured within an iterative loop. Since operators are loosely defined compute and memory transformations, our abstraction additionally allows them to be expressed with different \emph{execution policies} as a parameter to control synchronization behavior and parallelism. Much like the C++ standard library's execution policies~\cite{CPPREFERNCE}, these policies are unique types to allow for overloading of traversal and transformation operators to support parallelism and synchronization behaviors. Parallelism is supported by the work of an operator permitted to execute either in the invoking thread or in threads implicitly created by the operator's implementation. Synchronization behavior is supported by avoiding or introducing barriers on the invoking threads, based on the need of the timing model and the graph algorithm being implemented. These overloadings, now disambiguated with a unique type (\emph{execution policies}), allow for the operator's functionality to be identical, even as its underlying execution changes.



\subsection{Communication}
\label{sec:communication}

The two conventional communication models are shared-memory and message-passing. In a shared-memory model, all data is directly available to all processes, whereas in a message-passing model, data is made available through messages passed between processes~\cite{McCune:2015:TLA}. Expressing both models under the same framework can potentially allow for performance benefits in hierarchical distributed systems.
\textbf{The abstraction that enables support for multiple communication models is the use of frontiers with multiple underlying representations, which individually support shared-memory and message-passing models.}
When represented as an asynchronous queue~\cite{Chen:2021:AAT}, a frontier can communicate its elements using messages. When represented as a sparse vector or a dense bitmap~\cite{Wang:2017:GGG} stored in shared memory, its elements are directly available to all processes. With thoughtful design, regardless of the underlying representation, the top-level interface to query the frontier (or presence of an active vertex or edge) remains the same.

The communication model also goes hand-in-hand with the timing defined in the previous section. In a shared-memory communication model, performing bulk-synchronous operations on the frontiers with global synchronizations is a common, effective practice. However, depending on the size and workload imbalance of a frontier, an asynchronous execution model with message-passing to communicate the active working set can be more efficient.

\subsection{Vertex Programs and Push vs. Pull Traversals}
\label{sec:execution}

\textbf{The abstractions for supporting vertex or edge programs and traversal directions are exposed using three components:}
(1) the previously described operators, now with C++ lambda expressions~\cite{CPPREFERNCE} applied on the tuple \{source and destination vertices, and their corresponding edge\}, for every traversal or transformation (Section~\ref{sec:parallel-operators});
(2) the frontier type, expressed as either a set of active vertices or a set of active edges, which allows for both edge and vertex-centric programs~\cite{Wang:2017:GGG}; and
(3) the graph data structure stored as the original representation and the transposed representation, the former for push traversals and the latter for pull traversals, at the cost of memory space.

\subsection{Partitioning Schemes}
\label{sec:partitioning}

In our framework, partitioning space is largely left unexplored and is work in progress. However, since parts of our graph abstraction allow for multiple underlying representations, partitioned graphs could also simply be expressed as another such representation. Inheriting a partitioned graph within our framework would imply that when the top-level graph data structure is queried, the APIs will need to support the use of the corresponding partitioned sub-graph to return the result of a query.

\begin{table*} 
    \centering 
    \footnotesize
    \makebox[\textwidth]{
        \begin{tabular}{lllll}
        \toprule
        TLAV Pillars    & Models Captured                                                          & Abstraction                                                                              & Mechanism                                                                                                            & \begin{tabular}[c]{@{}l@{}}Models Ignored\\ (not captured)\end{tabular}                   \\
        \midrule
        Timing          & \begin{tabular}[c]{@{}l@{}}Bulk-Synchronous,\\ Asynchronous\end{tabular} & \begin{tabular}[c]{@{}l@{}}Operators,\\ Loop structure\end{tabular}                                                                                & Execution policies                                                                                                               &                                                                                     \\
        \midrule
        Communication   & \begin{tabular}[c]{@{}l@{}}Shared-Memory,\\ Message Passing\end{tabular} & \begin{tabular}[c]{@{}l@{}}Graph and Frontier\\ Representations\end{tabular}             & \begin{tabular}[c]{@{}l@{}}Queue-based (messages) or \\ bitmap, sparse frontiers\end{tabular}                                    & Active Messages                                                                           \\
        \midrule
        Execution Model & \begin{tabular}[c]{@{}l@{}}Vertex Programs,\\ Push vs. Pull\end{tabular} & \begin{tabular}[c]{@{}l@{}}Operators, Frontiers and\\ Graph Representations\end{tabular} & \begin{tabular}[c]{@{}l@{}}Vertex/edge-centric frontiers and\\ compressed sparse row/column\\ graph representations\end{tabular} &                                                                                           \\
        \midrule
        Partitioning    & Heuristics (Mostly Unexplored)                                           & \begin{tabular}[c]{@{}l@{}}Graph and Frontier\\ Representations\end{tabular}             & Random partitioning, METIS\cite{Karypis:1998:FHQ}                                                                                                       & \begin{tabular}[c]{@{}l@{}}Streaming, Vertex Cuts, \\ Dynamic Repartitioning\end{tabular} \\
        \bottomrule
        \end{tabular}
    }
    \caption{Summary of what models are captured within the four pillars of TLAV~\cite{McCune:2015:TLA} by our abstraction, and the corresponding element within the abstraction that implements the captured models. \label{tab:summary}}
\end{table*}

\section{Implementation}
\label{sec:implementation}
In this section, we show a vision of how to implement a graph framework  using modern C++ that captures a wide range of models described in the TLAV survey paper. Due to space constraints, we will only show the implementation of one algorithm (single-source shortest path), implemented with a bulk-synchronous timing model and shared-memory communication scheme in mind. Although we show a very limited scope of TLAV's four pillars, we highlight where the abstraction extends to support asynchrony, message-passing, and push- and pull-based graph processing.

\subsection{Graph representations}
\label{sec:representations}
The duality of graphs and sparse matrices can be exploited even in the native-graph approach for graph analytics. The underlying graph data structure can be expressed using common sparse matrix formats such as compressed-sparse row (CSR), compressed-sparse column (CSC), or an adjacency list. Sparse-matrix formats make for great graph representations due to the sparsity inherent in many large graphs and their ability to store such graphs in a compressed space. Listing~\ref{lst:graph-representation} shows one example of a graph internally represented as a CSR matrix, but queried with a graph-focused API\@.

\begin{tcolorbox}[title=++Push vs.\ Pull Traversal]
    \footnotesize{Our abstraction encompasses  the ability to inherit and retain multiple underlying data structures for a single graph at the same time. We can leverage multiple representations profitably; for instance, storing both CSR and CSC graph representations enables traversal models that support both push and pull.
    }
\end{tcolorbox}

\begin{listing}[h]
    \caption{\footnotesize{An example of requesting data from a sparse-matrix representation (CSR) with a native-graph API\@.}\label{lst:graph-representation}}
    \begin{minted}[
        fontsize=\footnotesize,
        autogobble, breaklines]{c++}
        // Compressed-Sparse Row (CSR) matrix.
        struct csr_t {
            int rows, cols;
            std::vector<int> row_offsets, column_indices;
            std::vector<float> values;
        };
        // In our framework, we rely on variadic inheritance
        // to support multiple underlying data structures.
        struct graph_t : public csr_t {
            // Get edge weight for a given edge.
            float get_edge_weight(int const& e) {
                return values[e];
            }
        };
    \end{minted}
\end{listing}

\subsection{Frontier representations}
\label{sec:frontier-representations}
Like graphs, frontiers can be represented with many different underlying representations. A sparse frontier can be simply represented as a vector of active vertex or edge indices. A dense frontier can be represented as a boolean array, where each element is true only if the corresponding vertex or edge is active. Depending on the scheduling and communication model, these frontier representations can be partitioned or be streamed to the compute units for processing. For a bulk-synchronous programming model with shared memory communication, the frontier can simply be stored in the shared memory (like the graph), and each element within the frontier can be processed in parallel. After each processing step, the synchronization step can be performed based on the operator's execution policy. In Listing~\ref{lst:frontier}, we show one example of how a sparse frontier can be implemented with a \mintinline{cpp}{std::vector} as its underlying data structure.

\begin{tcolorbox}[title=++Asynchrony and Message-Passing]
    \footnotesize{Based on Chen et al, we experiment with an asynchronous queue as an underlying structure to represent the frontier to allow for asynchrony and message-passing~\cite{Chen:2021:AAT}.}
\end{tcolorbox}

\begin{listing}[h]
    \caption{\footnotesize{Sparse frontier of active vertices represented as a simple vector from the C++ standard library.}\label{lst:frontier}}
    \begin{minted}[
        fontsize=\footnotesize,
        autogobble, breaklines]{c++}
        struct frontier_t {
            // Underlying representation of a frontier.
            std::vector<int> active_vertices;
            // Get the number of active vertices.
            int size() {
                return active_vertices.size();
            }
            // Get the active vertex at a given index.
            int get_active_vertex(int const& i) {
                return active_vertices[i];
            }
            // Add a vertex to the frontier.
            void add_vertex(int const& v) {
                active_vertices.push_back(v);
            }
        };
    \end{minted}
\end{listing}

\subsection{Parallel Operators}
\label{sec:parallel-operators}
A high-performance graph analytics implementation relies on efficient parallel operators that transform, expand, or contract the frontiers or graphs. This is where the bulk of optimizations can be introduced, such as utilizing data parallelism and load balancing. In Listing~\ref{lst:operators} we show an example of how a parallel operator performing a traversal (frontier expansion) can be expressed using modern C++, implemented on a BSP model with data stored in the shared memory space.

\begin{tcolorbox}[title=++Timing Model and Parallelism]
    \footnotesize{We extend our operators to support execution policies that overload the operator implementations to allow for parallel synchronous (\mintinline{cpp}{execution::par}) and asynchronous (\mintinline{cpp}{execution::par_nosync}) models.}
\end{tcolorbox}

\begin{listing}[h]
    \caption{\footnotesize{Synchronous parallel neighbor-expand, an operator derived from a traditional textbook graph algorithm~\cite{CLRS:2005:ITA} implemented using modern C++. Note, due to the limitations in C++20's execution policies,
    \john{is the previous phrase correct? i reworded it} \osama{yes!}
    the following version simply uses a parallel synchronous \mintinline{cpp}{std::for_each}; however, that can be replaced with parallel threads with no explicit barriers.
    }\label{lst:operators}}
    \begin{minted}[
        fontsize=\footnotesize,
        autogobble, breaklines, obeytabs=false, tabsize=1]{c++}
        #include <algorithm>
        #include <execution>
        #include <mutex>
        // Neighbor-expand implemented in C++20.
        template<typename expand_cond_t, typename policy_t,
            std::enable_if_t<
                !std::is_same_v<policy_t,
                decltype(execution::par_nosync)>, int> = 0>
        frontier_t neighbors_expand(
            policy_t execution_policy,
            graph_t& g, frontier_t& f,
            expand_cond_t condition) {
            std::mutex m;
            frontier_t output;
            auto expand = [=] (int const& v) {
                // For all edges of vertex v.
                for (auto e : g.get_edges(v)) {
                     auto n = g.get_dest_vertex(e);
                     auto w = g.get_edge_weight(e);
                     // If expand condition is
                     // true, add the neighbor into
                     // the output frontier.
                    if (condition(v, n, e, w)) {
                        std::lock_guard<std::mutex>
                            guard(m);
                        output.add_vertex(n);
                    }
                }
            };
            // For all active vertices in the
            // frontier, process in parallel.
            std::for_each(execution_policy,
                f.active_vertices.begin(),
                f.active_vertices.end(),
                expand);
            // Synchronized here and return output.
            return output;
        }
        // An alternative asynchronous version (par_nosync
        // is true) could launch parallel C++ threads and
        // avoid the synchronization entirely.
        template<typename expand_cond_t, typename policy_t,
            std::enable_if_t<
                std::is_same_v<policy_t,
                decltype(execution::par_nosync)>, int> = 0>
        frontier_t neighbors_expand(/*...*/) {/*...*/}
    \end{minted}
\end{listing}

\subsection{Example: Parallel Native-Graph SSSP}
\label{sec:example-parallel-native-graph-sssp}
To illustrate the abstraction, we show a simple example of a parallel SSSP algorithm implemented using the native-graph API, building on the previously described essential components. In Listing~\ref{lst:sssp}, we organize the algorithm in a BSP iterative while-loop with a convergence condition to schedule the completion. The key insights of the provided example are the use of (1) C++ lambda expressions to define a vertex program, (2) the \mintinline{cpp}{neighbor_expand} operator to perform the push-based traversal, and (3) \mintinline{cpp}{std::execution::par} to define the parallel execution policy.

\begin{listing}[h]
  \caption{\footnotesize{Parallel Single-Source Shortest Paths (SSSP) implemented in C++ using key components of native-graph graph analytics. Complete code:\\
      \url{https://github.com/gunrock/essentials-cpp}
      \john{Does the ``for'' below need braces or at least indentation after it?} \osama{Yes, it needed indentation.}
    }\label{lst:sssp}}
    \begin{minted}[
        fontsize=\footnotesize,
        autogobble, breaklines, obeytabs=false, tabsize=1]{c++}
        std::vector<float> sssp(
            graph_t& g, int const& source) {
            // Initialize data.
            std::vector<float> dist(g.get_num_vertices(),
                std::numeric_limits<float>::max());
            dist[source] = 0;
            frontier_t f; 
            f.add_vertex(source);
            while(f.size() != 0) {  // Main-loop.
                // Expand the frontier.
                f = neighbors_expand(
                    std::execution::par, g, f,
                    // User-defined condition for SSSP.
                    [=](int const& src,         // source
                        int const& dst,         // dest
                        int const& edge,        // edge
                        float const& weight) {  // weight
                        float new_d = dist[src] + weight;
                        // atomic::min atomically updates
                        // the distances vector at dst with
                        // the minimum of new_d or its
                        // current value, then returns the
                        // old value. (eq: mutex updates)
                        float curr_d =
                            atomic::min(&dist[dst], new_d);
                        return new_d < curr_d;
                    });
            }
            return dist;
        }
    \end{minted}
\end{listing}

\section{A Look Ahead}
As future work, we wish to explore many of TLAV's design decisions under a single framework targetting a wide-range of graph algorithms. Given the proposed abstraction, we make available ``essentials'', a graph library that targets GPUs:\\
\url{https://github.com/gunrock/essentials}.


\bibliographystyle{IEEEtran}
\bibliography{bib/graph,bib/owens,bib/algorithms,essentials}

\end{document}